\documentclass{article}
\begin{document}

\begin{center}
{\bf \large Advantages and Disadvantages of Layer Growth Model in Which Particles Maximize Number of Lateral Bonds\footnote{ Paper presented on 7-th Seminar on Thin Solid Films Structures and Surfaces (15-18 Sep. 1999, Kazimierz Dolny, Poland).}}

\bigskip

K. Malarz\footnote{ Electronic address: malarz@agh.edu.pl}

\bigskip

Department of Theoretical \& Computational Physics,
Faculty of Physics \& Nuclear Techniques,
University of Mining \& Metallurgy (AGH)\\
al. Mickiewicza 30, PL-30059 Krak\'ow, Poland.
\end{center}

\abstract{In this paper we would like give a short review of our recent works and current investigation in modeling of solid film surfaces growth based on computer Monte Carlo simulations.
We also discuss effect of some simplification in growth rules on speed-up of computations, and the time evolution of surface morphology (for example its roughness or anisotropy).
We obtained qualitative agreement of computer simulation with experimental results for homoepitaxially grown simple metal films.}

\bigskip


\section{Introduction}

In the last 40 years history of surface growth science, computer simulations were particularly useful and important for better understanding of modes and mechanisms of growth.
Great number of simulations is based on so-called solid-on-solid (SOS) model approximation, when each particle seats on the top of the others and film surface may be described by a single-valued function of its height $h(\mathbf{r})$ in planar substrate coordinate $\mathbf{r}$ (see [1-3] for review).
In this paper, we present a short review of our recent works [4-6] as well as current investigation [7,8] and oncoming ones [9].
In all these papers we concentrate on films growth on lattice with simple cubic symmetry.

	Section 2 is devoted to presentation of the model.
In Section 3 we present results which prove that a considerable speed-up of computations may be achieved at cost of some assumed simplification of growth rules and particle behaviors in computer simulation.
Then influence of model control parameters on surface morphology and its dynamics is presented in Section 3 and summarized in Section 4.

\section{Model}

Presented here results of simulations are based on simple SOS model of epitaxial thin films growth [4-8] based on Wolf-Villain model [10], in which after randomly chosen place of a particles initial contact with the surface, they relax to the site offering the strongest bonding (i.e. they maximize the number of particle-particle lateral bonds (PPLB)).
The place of particle settling is probed among site of initial deposition plus their four nearest-neighbors.
Then each particle repeats relaxation $L_{dif}$ times and so $L_{dif}$  may be considered as a range of diffusion.

	This simple and deterministic rules of growth [10] are generalized in this paper by accounting on:
\begin{itemize}
\item strengths of horizontal bonds ($J_x$, $J_y$) and vertical bonds ($S_x$, $S_y$) of PPLB in particle level layer and one layer below, respectively,
\item diffusion barrier ($V_x$, $V_y$), which restricts migration and stops it for infinitely large values of $V$,
\item temperature of substrate $T$. 
\end{itemize}

	The temperature $T$ entries calculation by the Boltzmann factor $\exp(-E/kT)$ for probability $p(E)$ of picking a site of energy $E$, which replaces the deterministic rule of fixing the particle at site of minimum $E$.
Desorption process is neglected becouse typical desorption energies for metals are of the order of several eV.
	With such modification we may grow anisotropic surfaces for model control parameters ($S_x\ne S_y$, $J_x \ne J_y$, and/or $V_x \ne V_y$) [6,8].
The diffusion barrier $V$ reduces probability of movement by factor $\exp(V_x/kT)$ or $\exp(V_y/kT)$ for considered direction $x$ or $y$ of migration. 
$V$ values are always positive, while negative $J$ and $S$ correspond to the assumed tendency to maximize PPLB.
Additionally, the presence of $S$ bonds to particles below in particle energy may play role of the Ehrlich-Schwoebel barrier near the film edge [11,12].
The $S$ term makes it difficult to reach the step edge and thus limiting horizontal growth of a cluster of atoms [5,13].

\section{Results of Simulations}
	The simulation were carried out on $L\times L$ large square lattice with periodic boundary conditions to minimize influence of boundary effects.
The particles are represented as unit cubes which can occupy only discrete position in (2+1)-D space.

\subsection{Diffusion Range}
	Firstly, we evaluate speed-up of computation when Monte Carlo simulations based on Arrhenius dynamics are reduced to oversimplified random deposition (RD) model through intermediate stages.
If $t_d$ is computing time to deposite a particle at the first contact site somewhere on the surface, then we need at least $t_{RD} = N\cdot t_d$ time, where $N$ is total number of the deposited particles, to complete simulation.
The RD model correspond to $L_{dif} = 0$. 
Increasing $L_{dif} = 1$ results in changing model to local diffusion with single step relaxation and increase time of computation by the additional time $t_r$ necessary for relaxation procedure: $t_{SSR} = N\cdot t_d+N\cdot t_r$.
Case $L_{dif} > 1$ corresponds to the local diffusion model when simulation time may be evaluated as $t_{LD} = N\cdot t_d+N\cdot L_{dif}\cdot t_r$.
Finally, for full-diffusion model ($L_{dif}\to\infty$) we have $t_{FD} = N\cdot t_d+(1+2+\dots+(N-1)+N)\cdot t_r$.
Note, that almost deterministic rules of growth speed-up computation on a ratio $k = t_{FD}/t_{LD} \approx (N+1)/2L_{dif} \approx \theta L^2/2L_{dif}$, where the total number of particles $N=\theta L^2$ is given by lattice lateral size $L$ and film thickness/substrate coverage $\theta$.
And --- as we shall show later --- for such simplification we may loose some significant film properties and/or misinterpret particles dynamics.

\begin{table}
\caption{Influence of bonding energy $J$ on roughness saturation level $\sigma_{sat}$ for large diffusion range $L_{dif} \gg L$.
$L_{opt}$ gives a rough estimate of the diffusion range for which saturation of $\sigma$ starts.
$\theta$ =  10 ML, $V = S = 0$.}
\begin{center}
\begin{tabular}{rrrrrrrrrrr}
\hline
$J$ & 0.00& $-0.05$& $-0.10$& $-0.25$& $-0.50$& $-0.75$& $-1.00$& $-5.00$& $-20.0$\\
\hline
$L_{opt}$ & 0& 1& 5& $\approx$50& $\approx$100& $\approx$200& $\approx$300& $\approx$500& $>5000$\\
$\sigma_{sat}$ & 3.16& 2.83& 2.53& 1.83& 1.30& 1.10& 1.00& 0.50& 0.36\\
\hline
\end{tabular}
\end{center}
\end{table}

	The range of diffusion $L_{dif}$ influences surface roughness measured by standard surface width $\sigma$, defined as the root-mean-square of surface heights.
We found that increasing of diffusion range make surface smoother, however some saturation level $\sigma_{sat}(L_{dif}\to\infty)>0$ is observed.
The saturation level depends on strong of attractive forces among particles which is governed by $J$ [5].
The results are collected in Tab. 1.
Note, that effective diffusion range $L_{eff}$ may be also controlled by substrate temperature $T$ as we show later in Section 3.4.

\subsection{Growth Dynamics}
	The roughness $\sigma$ for film surfaces growing by deposition often obeys Family-Vicsek scaling law: $\sigma\propto L^\alpha\cdot f(\theta /L^z)$, where scaling function is given by $f(x)\to x^\beta$ for $x\ll 1$ and $f(x)\to 1$ for $x\gg 1$ [14,15].
For substrate low coverage ($\theta < 1$ ML) also third regime is observed with growth exponent $\beta = \beta_{RD} =1/2$, similar to simple RD [5,8,15,16].
Deviation from scaling hypotheses may be also observed when particles are allowed to climb on the top of other ones which results an unstable growth when surface width growth infinitely with increasing of coverage $\theta$ [5,17,18].
This may be considered as infinite value of roughness exponent ($\alpha\to\infty$).

When also anisotropy in growth is included (e.g. for Ag/Ag(110) or Cu/Cu(110)), then we observe power law for roughness $\sigma\propto\theta^\beta$ for small $\theta$ [8] (model parameters for Cu and Ag are taken after [19]).
Before reaching $\theta = 10^4$ ML we still do not observe roughness saturation which may indicate that particles jumps to the higher levels may be not essential for presence of unstable growth ($\alpha\to\infty$) as it was earlier suggested in [5,18].
In both cases, either stable or unstable growth, surface roughness $\sigma$ increases initially as $\theta^\beta$.
The growth exponent $\beta$ may also depend on substrate crystallographic orientation as was shown by Elliot {\it et al.} for Ag(111) and Ag(100) [20].
This can be explained only within models that allow for anisotropy.

Family-Vicsek law fails for inhomogeneous multilayerd $a/b/a$-like sandwiches as a result of effective changes of growth rules --- or more precisely --- strength of interaction during deposition [7].

\subsection{Surface Morphology Anisotropy}
	As we mentioned in Section 2, the anisotropy may originate from either anisotropy in bonds ($S_x \ne S_y$, $J_x \ne J_y$) or anisotropy in barriers $V_x \ne V_y$ [6,8].
Then anisotropy may be quantitatively measured from the difference in the height-height correlation function $G(\mathbf{s}) \equiv \langle h(\mathbf{r}+\mathbf{s})\cdot h(\mathbf{r}) \rangle - \langle h(\mathbf{r}) \rangle ^2$ in two $x$ and $y$ directions, where $\langle\cdots\rangle$ denotes spatial average over all sites $\mathbf{r}$.
Correlation $G(1,0)$ are positive for terrace-like, smooth surfaces (in $x$- direction) while its negative values correspond to spiky and rough structures.
Thus $G(1,0)-G(0,1)$ normalized to $\sigma^2=G(0,0)$ may be used as a quantitative measure of anisotropy $\varepsilon\equiv [G(1,0)-G(0,1)]/G(0,0)$ [6] and may be competitive to a direct pictures of the surface morphology obtained from simulations [8].
We found that anisotropy parameter $\varepsilon$ depends on the ratio of $|J_x-J_y|/|J_x+J_y|$ and not on the simple difference $|J_x-J_y|$ only.
As it may be expected, the anisotropy vanishes for very large diffusion barriers $V_x = V_y\to\infty$, when RD results are reproduced.
When the first monolayer is completed, the anisotropy parameter $\varepsilon$ seems to become insensitive to increasing coverage $\theta$ [6].

\subsection{Influence of Substrate Temperature on Film Properties}
We expect that increasing temperature should make the surface smoother since energetic barriers are easier to overcome and also surface bonds may be broken more easily. 
The effect of temperature increase is presented in Tab. 2.
Model parameters are taken after [13] where simulations of Pt on Pt(111) behaviors were studied.

	Dependence of effective diffusion range with substrate temperature may also be easily predicted; for low temperatures particles freeze at the place of deposition and migration is stopped ($T\to0 \Rightarrow L_{eff} = 0$).
Then we reproduce RD results with Poisson distribution of film heights.
On contrary, hot substrate make particles mobile ($T\to\infty \Rightarrow L_{eff}\to\infty$) and particles stop only if diffusion range is reached in the number of subsequent jumps [8].
The quantity introduced to describe particle mobility is number $M$ of particles which are still unsettled on the surface when the diffusion range is reached, that is the number $M$ of particles which changed their position during the last step ($L_{dif}$) any iteration cycle.
$M$ is increasing function of substrate temperature $T$, as expected (see Tab. 3).
Note, that the increase in $L_{dif}$ plays similar role as an increase in substrate temperature $T$ for stochastic deposition processes.

\begin{table}
\caption{Dependence of roughness $\sigma$ on substrate temperature $T$ for Pt on Pt(111) diffusion.
Model control parameters are after Ref. 13.}
\begin{center}
\begin{tabular}{rrrrrr}
\hline
$T$ [K]& 300& 600& 900& 1200& 1500\\
\hline
$\sigma$ [ML]& 2.04& 1.87& 1.74& 1.60& 1.49\\
\hline
\end{tabular}
\end{center}
\end{table}

\begin{table}
\caption{Number $M$ of mobile particles after $L_{dif} =10^3$ jumps for homoepitaxial growth of Cu/Cu(110) and Ag/Ag(110).
The model parameters are after Ref. 19.}
\begin{center}
\begin{tabular}{rrrrrr}
\hline
$T$ [K]& 300& 600& 900& 1200& 1500\\
\hline
Cu/Cu(110)& 1& 50& 239& 913& 1348\\
Ag/Ag(110)& 0& 28& 147& 529& 1069\\
\hline
\end{tabular}
\end{center}
\end{table}

\begin{table}
\caption{Influence of substrate temperature $T$ on surface morphology anisotropy for Cu/Cu(110) and Ag/Ag(110) [8].
Model control parameters are taken after Ref. 19.}
\begin{center}
\begin{tabular}{rrrrrr}
\hline
$T$ [K]& 300& 600& 900& 1200& 1500\\
\hline
$\varepsilon$ (Cu)& 0.275& 0.732& 0.828& 0.535& 0.508\\
$\varepsilon$ (Ag)& 0.197& 0.563& 0.729& 0.585& 0.453\\
\hline
\end{tabular}
\end{center}
\end{table}

	Finally, we check how temperature influences surface morphology measured by $\varepsilon$  and applied to Cu/Cu(110) and Ag/Ag(110).
For absolute zero temperature ($T\to 0$) we recover RD results ($\varepsilon\to 0$).
With increasing temperature the anisotropy coefficient reaches maximum and then drops, so that, we again have RD results $\varepsilon = 0$ for $T\to\infty$.
In intermediate temperature surface morphology evolves from isolated small islands, through strongly anisotropy long string of atoms along dimmer rows to larger and more isotropic islands also across dimmer rows [8,19].

\section{Summary}
	The described model proposed in [4-8] allows for qualitatively correct description of thin solid films grown by molecular beam epitaxy where dominant physical processes responsible for film growth is surface diffusion.
We show that simplification of rules of growth --- when next particle incoming only when position of a previous one was fixed --- speeds up computation drastically particularly for large lattices and large coverage.
This is justified for not very large flux of the incoming particles.
Actually, bigger $L_{dif}$ also mimics smaller flux of the deposited particles.
The anisotropy in the model parameters allows to consider not only sc(100) symmetry surfaces, but also particles deposited on bcc(110) surfaces [8].

	From simulation we were able to reproduce data on thermal, time and structural evolution of surface morphology.
Usually the surface roughness measured by standard deviation of film height increases with time (or coverage) accord to power law [5,8].
Such behavior is characteristic for early stages of growth of self-affine surfaces (see [21] for review).

	We believe that it is important to study the fractal dimension for the (2+1)-D surfaces for better understanding of the growth phenomena.
We also intend to investigate more throughly the anisotropic growth [9].

\section*{Acknowledgments}
	I am very grateful to A.Z.Maksymowicz for fruitful discussions and scientific guidance.
Paper was partially supported by Polish Committee for Scientific Research (KBN) 
under grant 8 T11F 02616.
The calculations were carried out in ACK-CYFRONET- AGH.
The machine time is financed by KBN with grants KBN/C3840/AGH/008/1997, KBN/S2000/AGH/069/1998 and KBN/HP-K460-XP/AGH/069/1998.


\end{document}